\documentclass{svjour3}
\usepackage{times}
\usepackage{blindtext, graphicx}
\usepackage{cite}
\usepackage{colortbl}
\usepackage{tikz}
\usepackage{microtype}

\smartqed

\usepackage{balance}
\definecolor{Gray}{rgb}{0.88,1,1}
\definecolor{Gray}{gray}{0.85}
\definecolor{lightgray}{gray}{0.8}
\usepackage{subfig} 

\usepackage[linesnumbered,ruled,vlined]{algorithm2e}
\SetKwProg{Fn}{Function}{}{}

\usepackage{amsmath}

\usepackage[framed]{ntheorem}
\usepackage{framed}
\usepackage{tikz}
\usetikzlibrary{shadows}
\theoremclass{Lesson}
\theoremstyle{break}

\tikzstyle{thmbox} = [rectangle, rounded corners, draw=black,
fill=Gray!20,  drop shadow={fill=black, opacity=1}]

\newshadedtheorem{lesson}{Finding}

\begin{document}

    \title{ Finding Better Active Learners   for Faster Literature Reviews
}


\author{Zhe Yu         \and
        Nicholas A. Kraft \and 
        Tim Menzies
}


\institute{Zhe Yu \at
              Department of Computer Science, North Carolina State University, Raleigh, NC, USA \\
              \email{zyu9@ncsu.edu}           
           \and
           Nicholas A. Kraft \at
              ABB Corporate Research, Raleigh, NC, USA\\
              \email{nicholas.a.kraft@us.abb.com}
            \and
           Tim Menzies \at
              Department of Computer Science, North Carolina State University, Raleigh, NC, USA \\
              \email{tim.menzies@gmail.com}
}


\maketitle

\begin{abstract}
  
Literature reviews can be time-consuming and tedious to complete.
By cataloging and refactoring three state-of-the-art active learning techniques from evidence-based medicine and legal electronic discovery, this paper finds and implements FASTREAD, a  faster technique for  studying a large corpus of documents, combining and parametrizing the most efficient active learning algorithms. This paper assesses FASTREAD using   datasets generated from existing SE literature reviews (Hall, Wahono, Radjenovi{\'c}, Kitchenham et al.).
Compared to manual methods, FASTREAD lets 
researchers find 95\% relevant studies after reviewing an order of magnitude fewer papers. Compared to  other state-of-the-art automatic methods,  FASTREAD 
reviews 20-50\% fewer studies while finding same number of relevant primary studies in a systematic literature review.


\end{abstract}

\section{Errarta}
\label{sect: Errarta}

This version contains updated results after fixing issues in previous datasets. The new results make the same overall conclusion as before; i.e. FASTREAD is best. The authors thank Chandramouli Shama Sastry for pointing out those issues in the datasets. The updated datasets are available at {\em https://doi.org/10.5281/zenodo.1162952}.

\section{Introduction}
\label{sect: Introduction}

When conducting a literature review in
 software engineering, it is common practice~\cite{kitchenham2013systematic} to
 conduct  a {\em primary study selection} where
 a large number of potentially
 relevant papers, collected via some initial query (e.g. keyword search in Google Scholar), are manually reviewed and assessed for relevance. To reduce the effort associated with conducting such tedious and time-consuming
{\em linear manual reviews}, researchers in  the fields of   evidence-based medicine~\cite{paynter2016epc,wallace2010semi,wallace2010active} and  
electronic discovery~\cite{cormack2014evaluation,cormack2015autonomy} have developed {\em active learning}
methods that can build an automatic classifier
that prunes away irrelevant papers  using feedback from users.

In this paper we investigate whether there are any insights from  that related
work that can reduce the effort associated with literature reviews
in software engineering (SE). Specifically, in this paper we:
\begin{itemize}
\item
Review those active learning methods~\cite{paynter2016epc,wallace2010semi,wallace2010active,cormack2014evaluation,cormack2015autonomy} and find that in evidence-based medicine and legal electronic discovery, there are three widely recognized state-of-the-art active learning methods~\cite{cormack2014evaluation,wallace2010semi,miwa2014reducing}.
\item
Analyze those three active learning methods and find that they are each assembled from lower-level techniques that address four questions: (1) when to start training, (2) which study to query next, (3) whether to stop training, and (4) how to balance the training data.
\item
Investigate 32 possible active learning approaches that represent different combinations of lower-level techniques to address the four questions.
\item
Evaluate the 32 active learning approaches using SE data and the evaluation criteria ``work saved over sampling at 95\% recall'' (WSS@95)~\cite{cohen2011performance}. 
\item
Discover that one of those 32 active learning approaches, which we   call FASTREAD, reduces the effort required to find relevant papers the most.
\end{itemize}
Based on that work, the   contributions and outcomes of this paper are:
\begin{enumerate}
\item
A cautionary tale that  verbatim  reuse of data mining methods
from other fields may not produce the best results for SE.
Specifically, we  show that supposed state-of-the-art methods
from other fields do not work best on SE data.
\item
A case study showing the value  of refactoring 
and recombining data mining methods.  The FASTREAD tool recommended by this
paper was constructed via such refactoring.
\item
 A demonstration that  FASTREAD is a new highwater mark in
 reducing the effort associated with  primary study selection in SE literature reviews.
 \item
 A open source workbench that allows for the fast evaluation of FASTREAD,
 or any other technology assisted reading method. See https://github.com/fastread/src. 
 \item
Four new data sets that enable extensive evaluation of FASTREAD or other methods. The creation and distribution of these data sets is an important contribution, because prior to this study, it was  very difficult to obtain even one such data set. 
\end{enumerate}
The rest of this paper offers background notes on the problem of reading technical documents and on how that problem has been solved
in other fields. We then refactor those solution into 32
candidate solutions, which we asses using  prominent  published SE literature reviews.  Using that data,
we ask and answer the following three research questions:

\begin{itemize}

\item
{\bf RQ1: Can active learning techniques reduce effort in primary study selection?}  We find that using FASTREAD,  after reviewing a few hundred papers,  it is possible
to find   95\% of the relevant papers found by a linear
manual review of thousands of papers.   
\item
{\bf RQ2: Should we just adopt the state-of-the-art treatments from other fields?} Our results show that better active learners
for SE can be build by  mixing and matching methods
from the state-of-the-art in other fields.
\item
{\bf RQ3: How much effort can FASTREAD, our new state-of-the-art method for primary study selection, save in an SLR?} We show that FASTREAD can reduce more than 40\% of the effort associated with 
the primary selection study phase of a literature review
while retrieving 95\% of the relevant  studies. 

\end{itemize}

\begin{figure}[!b]
    \centering
    \subfloat[Most Difficult Aspects of SLR Process.]
    {
        \includegraphics[width=0.48\linewidth]{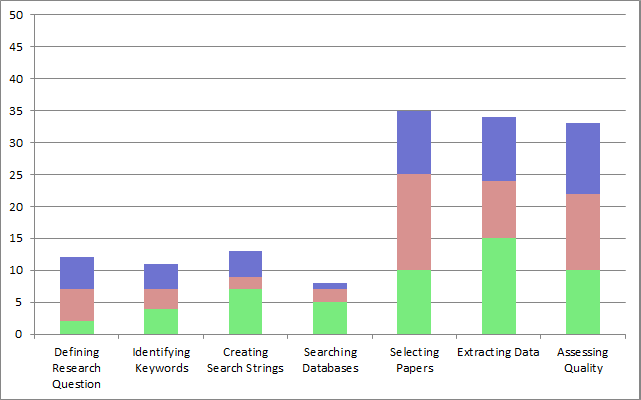}
        \label{fig: difficult}
    }
    \subfloat[Most Time Consuming Aspects of SLR Process.]
    {
        \includegraphics[width=0.48\linewidth]{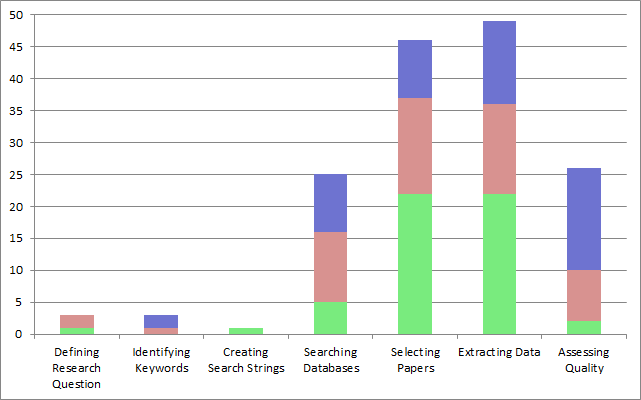}
        \label{fig: time}
    }    
    \caption{Data collected from surveys to SLR authors~\cite{carver2013identifying}. {\setlength{\fboxsep}{1pt}\colorbox{green!40}{green}}, {\setlength{\fboxsep}{1pt}\colorbox{red!30}{red}}, and {\setlength{\fboxsep}{1pt}\colorbox{blue!50}{blue}} show the most,   second most, and third most, respectively voted item.}
    \label{fig:barrier}
\end{figure}

\section{Background}
\label{sect: sect2}


Systematic Literature Reviews
(SLRs)  are a well established and widely
applied review method in Software Engineering since Kitchenham, Dyb{\aa}, and
J{\o}rgensen first adopted it to support evidence-based software engineering in
2004 and 2005~\cite{kitchenham2004evidence,1377125}. 
Researchers can get a
general idea of current activity in their field of interests by reading the SLR studies. Furthermore, a
deeper understanding of the topic may be gained by conducting an SLR.

An increasing number of SLRs has been conducted since the proposal and revision of the SLR guidelines in 2007~\cite{keele2007guidelines}. For example, there were 26 SLRs on IEEE Xplore during the year of 2005 and that
number has increased to 137, 199 for the years 2010, 2015 (respectively). Various scholars  suggest that an SLR is required before any research in Software
Engineering is conducted~\cite{keele2007guidelines}.
While this is certainly a good advice,
currently an SLR is
a large, time consuming and complex
task~\cite{hassler2016identification,hassler2014outcomes,carver2013identifying,bowes2012slurp}.

Cost reduction in SLRs is therefore an important topic and will benefit researchers in software engineering community.
Previously we have analyzed the costs of SLRs~\cite{hassler2014outcomes,carver2013identifying}. As shown in Fig.~\ref{fig:barrier}, primary study selection, which is noted as ``selecting papers'' in Fig.~\ref{fig:barrier}, is among the top three most difficult as well as time-consuming aspects in an SLR. Usually, reviewers need to evaluate thousands of studies trying to find dozens of them that are relevant to the
research questions based on their title, abstract, or full text~\cite{bowes2012slurp}. An extreme
example of this is where reviewers sourced over
3,000 studies, and only used 7 of them in their final review~\cite{bezerra2009systematic}.  In terms of actual
cost, Malheiros has documented that it requires 3 hours for one reviewer to review 100 studies~\cite{malheiros2007visual}.  This implies that it is a month's work for one graduate student to review 3000 studies or three months' work to review 9000 studies. The cost associated with primary study selection has become a serious problem and will continue to grow in the near future as the population of candidates for primary studies increases dramatically. In this paper, we focus on reducing cost in primary study selection only. Our prioritization on the cost reductions of primary study selection is not to discount the effort associated with other parts of the SLR process. Indeed, one of our considerations is that there are already tools to support other parts of the SLR process, as well as different techniques to facilitate primary study selection. All these techniques (such as Quasi-Gold Standard based search~\cite{zhang2011empirical,zhang2011identifying}, visual text mining~\cite{Felizardo:2014:VAA:2601248.2601252,felizardo2012visual,felizardo2010approach,malheiros2007visual}, and snowballing~\cite{wohlin2014guidelines,jalali2012systematic}) are compatible with works in this paper and a better performance is expected when applied together. This leads to a direction of future work in which the best setting to integrate different techniques will be explored.

There are three main aspects in primary study selection: \textbf{(1)} retrieving initial list of primary studies, \textbf{(2)} excluding irrelevant studies, \textbf{(3)} including missing studies. We focus on excluding irrelevant studies because \textbf{a)} there already exists techniques and tools to facilitate \textbf{(1)} and \textbf{(3)} such as Snowballing~\cite{jalali2012systematic} and StArt~\cite{hernandes2012using}; \textbf{b)} the performance of excluding irrelevant studies can be evaluated using existing SLR publications.

\subsection{Related Work}
\label{sect: Background}

\subsubsection{Software Engineering Tools}

In recent years, various tools have been developed to facilitate SLRs in the software engineering community, as summarized in the associated SLRs~\cite{marshall2015tools,marshall2014tools,marshall2013tools}. These tools aim at providing support for protocol development~\cite{Molleri:2015:SWA:2745802.2745825,fernandez2010slr,hernandes2012using}, automated search~\cite{Molleri:2015:SWA:2745802.2745825,hernandes2012using}, primary study selection~\cite{Molleri:2015:SWA:2745802.2745825,hernandes2012using,fernandez2010slr,bowes2012slurp}, quality assessment~\cite{fernandez2010slr,bowes2012slurp,Molleri:2015:SWA:2745802.2745825}, data extraction and validation~\cite{Molleri:2015:SWA:2745802.2745825,hernandes2012using,fernandez2010slr,bowes2012slurp}, data synthesis~\cite{Molleri:2015:SWA:2745802.2745825,hernandes2012using,fernandez2010slr,bowes2012slurp},
and report write up~\cite{Molleri:2015:SWA:2745802.2745825,hernandes2012using,fernandez2010slr,bowes2012slurp}. It is extremely helpful to have a tool for managing the whole SLR process. However, the support for primary study selection using these tools is limited (e.g., to tasks such as assigning review jobs to multiple reviewers or to resolving disagreements).
Hence, we planned to introduce machine learning to assist primary study selection in SE SLRs but before this paper is published, Ros et al.~\cite{ros2017machine} has achieved this in June 2017. While Ros'17~\cite{ros2017machine} provided a wide range of techniques to support both search and selection, it has several limitations such as (a) not comparing against state-of-the-art techniques from other domains (which are approaches discussed later in Section~\ref{sect: Electronic Discovery} and~\ref{sect: Evidence-based Medicine}); (b) not considering any data balancing; (c) testing only on a single unpublished dataset.

Visual text mining (VTM) is a technique especially explored in Software Engineering community to support SLR. It is an unsupervised learning method which visualizes the relationship between candidate studies and helps the reviewer to make quick decisions. Malheiros et al.~\cite{malheiros2007visual} first applied VTM to support primary study selection in SLR. In their small-scale experiment (100 candidate studies, 31 of which are ``relevant''), VTM retrieves around 90\% of the ``relevant'' studies by spending about 30\% as much time as manual review. However, VTM requires some prior experience and knowledge of text mining and visualization techniques to use~\cite{bowes2012slurp}, and more case studies with large scale are needed to validate their results.

Snowballing is another technique attracting much attention in SE SLR research. Given the inherent relevance relationship between a study and its citations, it is of high probability for the citations of (used in backward snowballing) and the studies cite (used in forward snowballing) a known ``relevant'' study to also be ``relevant''~\cite{kitchenham2004evidence}. Jalali and Wohlin~\cite{jalali2012systematic,wohlin2014guidelines} applied backward snowballing to search for primary studies in SE SLRs and found comparably good result as database search. Felizardo et al.~\cite{felizardo2016using} and Wohlin~\cite{wohlin2016second} applied forward snowballing to update SE SLRs and greatly reduced the number studies need to be reviewed comparing to a database search. This paper does not use snowballing since, as mentioned by Wohlin~\cite{wohlin2014guidelines}, snowballing starts with an initial set of relevant papers.
FASTREAD's task is very different: we start with zero relevant papers.

\subsubsection{Legal Electronic Discovery Tools}
\label{sect: Electronic Discovery}

Electronic Discovery (e-discovery) is a part of civil litigation where one party (the producing party), offers up materials which are pertinent to a legal case~\cite{krishna2016bigse}. This involves a review task where the producing party need to retrieve every ``relevant'' document in their possession and turn them over to the requesting party. It is extremely important to reduce the review cost in e-discovery since in a common case, the producing party will need to retrieve thousands of ``relevant'' documents among millions of candidates. Technology-assisted review (TAR) is the technique to facilitate the review process. The objective of TAR is to find as many
of the ``relevant'' documents in a collection as possible, with reasonable cost~\cite{grossman2013}. Various machine learning algorithms have been studied in TAR. So far, in every controlled studies, continuous active learning (Cormack'14) has outperformed others~\cite{cormack2014evaluation,cormack2015autonomy}, which makes it the state-of-the-art method in legal electronic discovery. It has also been selected as a baseline method in the total recall track of TREC 2015~\cite{roegiest2015trec}. Details on continuous active learning are provided in Section~\ref{sect: Methods}. 



\subsubsection{Evidence-based Medicine Tools}
\label{sect: Evidence-based Medicine}

Systematic literature reviews were first adopted from evidence-based medicine in
2004~\cite{kitchenham2004evidence}. To facilitate citation screening (primary
study selection) in systematic review, many groups of researchers have investigated different types of machine learning algorithms and evaluation mechanisms~\cite{o2015using,paynter2016epc}. 

Cohen et al. first applied text mining techniques to support citation screening and developed several performance metrics (including WSS@95) for assessing the performance of different techniques in 2006~\cite{cohen2006reducing}. While the great contribution of introducing machine learning and text mining into citation screening as well as the proposed performance metrics of Cohen has been widely acknowledged~\cite{o2015using}, most of Cohen's work focused on supervised learning which does not utilize unlabeled data and relies on random sampling to obtain the sufficiently large training set~\cite{cohen2006reducing,cohen2006effective,cohen2010prospective,cohen2011performance}.

Wallace et al. conducted a series of studies
with machine learning techniques, especially active
learning~\cite{wallace2010semi,wallace2010active,wallace2011should,wallace2012deploying,wallace2013active,wallace2013modernizing,nguyen2015combining}. Wallace
first set up a baseline approach called ``patient active learning'' (Wallace'10) for machine learning assisted citation screening~\cite{wallace2010semi}. The performance of patient active learning is good enough (nearly 100\% of the ``relevant''
citations can be retrieved at half of the conventional review cost) to convince
systematic review conductors to adopt machine learning assisted citation
screening. Instead of improving this baseline method, Wallace then focused on other aspects of machine learning assisted citation screening such as introducing external expert knowledge~\cite{wallace2010active}, allocating review tasks to multiple experts~\cite{wallace2011should} or to crowdsourcing workers~\cite{nguyen2015combining}, and building a tool called abstrackr to provide overall support~\cite{wallace2012deploying}. Wallace's work on this topic is of exemplary high-impact and his core algorithm   (on simple expert screening),   is one of the most popular active learning techniques we have found in the evidence-based medical literature. That said, this technique has not been updated since 2010~\cite{wallace2010semi}. In this paper we are focused on the core active learning algorithm for cost minimization. Hence, we do not explore techniques such as Wallace's use of multiple experts (but in future work, we will explore this approach).

More recent work of Miwa et al. explored alternative data balancing and query strategy in 2014~\cite{miwa2014reducing} and proposed a new treatment of Certainty plus Weighting (Miwa'14). Instead of uncertainty sampling in patient active learning (Wallace'10), Miwa found that certainty sampling provides better results in clinical citation screening tasks. Similar conclusion for data balancing method as weighting relevant examples was found to be more effective than aggressive undersampling. Although not stated explicitly, Certainty plus Weighting keeps training until all ``relevant'' studies have been discovered, which differs from the stopping criteria of Wallace'10. Aside from the core algorithm, additional views from latent Dirichlet allocation (LDA) has been found to be potentially useful.

Other work related to machine learning assisted citation screening do not
utilize active learning. Pure supervised learning requires a sufficiently large training set, which leads to a huge review cost~\cite{cohen2006reducing,adeva2014automatic}. Semi-supervised learning~\cite{liu2016comparative} does not utilize the human reviewers' feedback for updating the model, which leads to a depreciated performance in a long run. As a result, the patient active learning proposed by Wallace et al.~\cite{wallace2010semi} and the Certainty plus Weighting approach by Miwa et al.~\cite{miwa2014reducing} are still considered to be the state-of-the-art method for citation screening in the scenario with no external knowledge and equally expensive reviewers. Details on these two approaches are provided in Section~\ref{sect: Methods}.

There are also existing tools to support study selection in systematic reviews, e.g. Abstrakr\footnote{http://abstrackr.cebm.brown.edu}~\cite{wallace2012deploying}, EPPI-Reviewer\footnote{http://eppi.ioe.ac.uk/cms/er4/}~\cite{thomas2010eppi}, Rayaan\footnote{http://rayyan.qcri.org/}~\cite{Ouzzani2016}. Useful features can be found in these tools such as a) Rayaan and EPPI-Reviewer: incorporated keyword search in screening; b) Rayaan and EPPI-Reviewer: deduplication; c) Rayaan and EPPI-Reviewer: define inclusion/exclusion criteria by terms; d) Abstrakr: user defined tags; e) all three: assign review tasks to multiple reviewers; f) all three: automatically extract data from PubMed. However, the active learning parts alone in these tools are depreciated. Under the condition that no additional feature (search, tags, define inclusion/exclusion terms) is used, we tried all three tools with one of our dataset-- Hall set (104 relevant in 8911 studies) and after reviewing 1000 studies, only 10 to 15 relevant ones were found, which was very close to a random sampling result without any learning. Since none of these tools are open-source, we cannot tell whether active learning is applied or how/when it is applied in each tool. This motivates us to develop an open source tool which focuses on active learning to support the primary study selection process. Details about our tool are presented in Section~\ref{sect: tool}.

\section{Technical Details}
\label{sect: Methods}

As mentioned in Section~\ref{sect: Background}, the existing state-of-the-art methods are Wallace'10~\cite{wallace2010semi} (patient active learning), Miwa'14~\cite{miwa2014reducing} (Certainty plus Weighting), and Cormack'14~\cite{cormack2014evaluation} (continuous active learning). All three state-of-the-art methods share the following common techniques.

\textbf{Support vector machines (SVM)} are a well-known and widely used classification technique. The idea behind is to map input data to a high-dimension feature space and then construct a linear decision plane in that feature space~\cite{cortes1995support}. Linear SVM~\cite{joachims2006training} has been proved to be a useful model in SE text mining~\cite{krishna2016bigse} and is applied in the state-of-the-art active learning methods of both evidence-based medicine and electronic discovery~\cite{miwa2014reducing,wallace2010semi,cormack2014evaluation}. One drawback of SVM is its poor interpretability as compared to classifiers like decision trees. However, SVM still fits here since the model itself is not important as long as it could provide a relative ranking of literature.

\textbf{Active learning} is a cost-aware machine learning algorithm where labels of training data can be acquired with certain costs. The key idea behind active learning is that a machine learning algorithm can perform better with less
training if it is allowed to choose the data from which it learns~\cite{settles2012active}. There are several scenarios active learning is applied to, such as membership query synthesis, stream-based selective sampling, and pool-based sampling~\cite{settles2010active}. There are also different query strategies of active learning, such as uncertainty sampling, query-by-committee, expected model change, expected error reduction, variance reduction, and density-weighted methods~\cite{settles2010active}. Here, we briefly introduce one scenario and two query strategies, which are used in our later experiments and discussions.

\begin{figure}[!t]
    \centering
    \subfloat[No data balancing]
    {
        \includegraphics[width=0.33\linewidth]{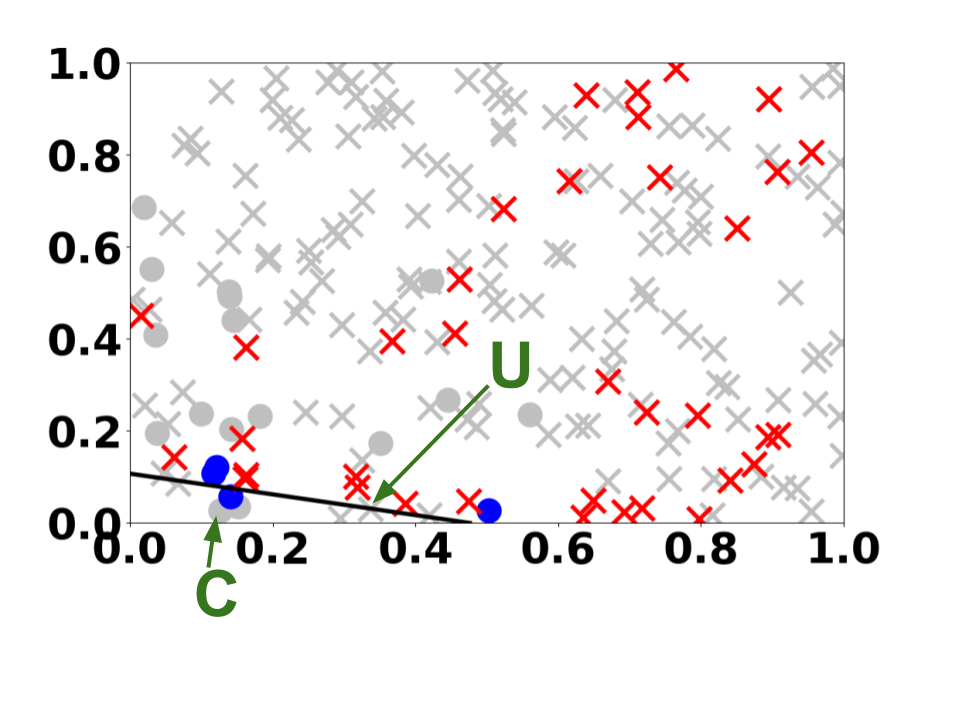}
        \label{fig:nobal}
    }
    \subfloat[With weighting]
    {
        \includegraphics[width=0.33\linewidth]{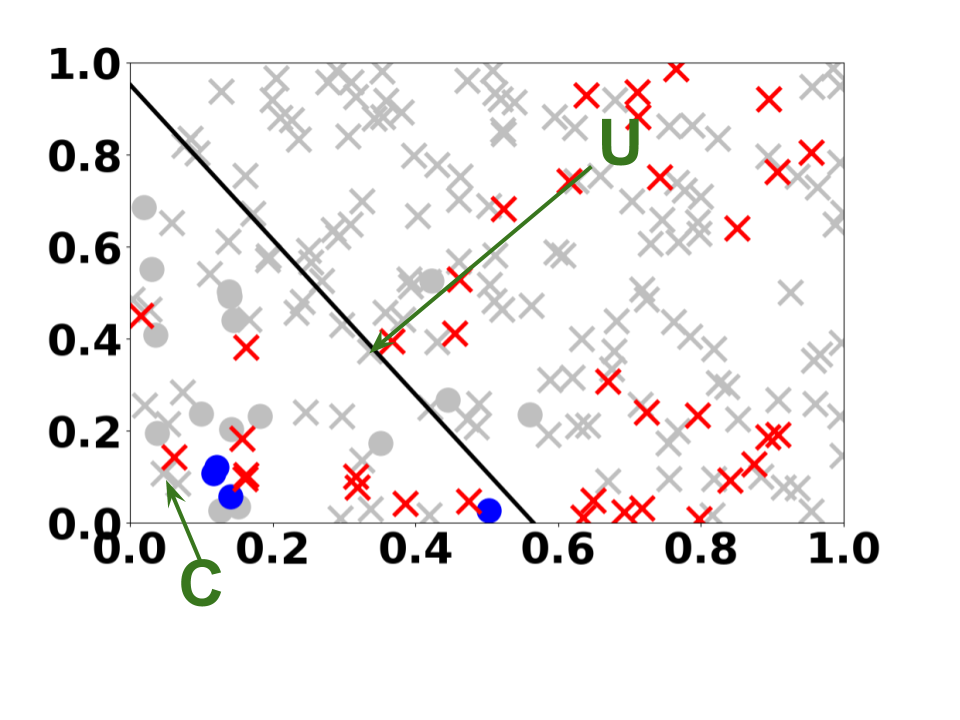}
        \label{fig:weighting}
    }
    \subfloat[With aggressive undersampling. ]
    {
        \includegraphics[width=0.33\linewidth]{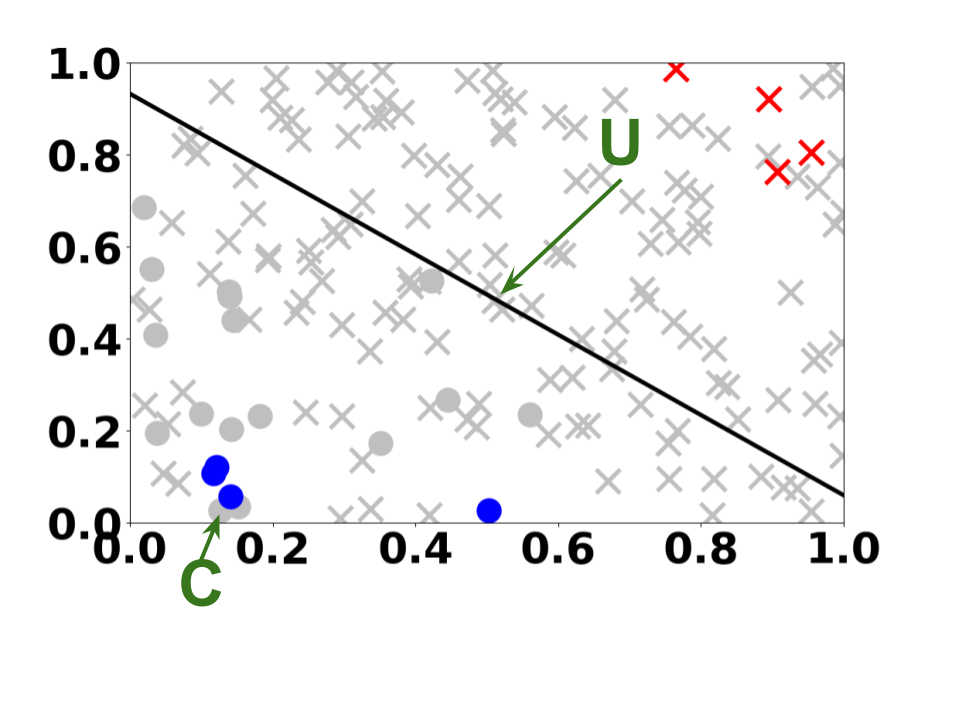}
        \label{fig:aggress}
    }
    \caption{Active learning with SVM and different data balancing techniques.}
    \label{fig:SVM}
\end{figure}

Figure~\ref{fig:SVM} shows a simple demonstration of an SVM active-learner.
For the sake of simplicity, this demonstration assumes that the data has    two features (shown in that figure as the horizontal and vertical axis). 
 In that figure, ``$O$'' is the minority class, ``relevant'' studies in SLR. ``$O$''s in blue are studies already identified as ``relevant'' (included) by human reviewers. ``$X$'' is the majority class, ``irrelevant'' studies in SLR. ``$X$''s in red are studies already identified as ``irrelevant'' (excluded) by human reviewers (note that in (c), some red ``$X$''s are removed from the training set by aggressive undersampling). Markers in gray are the unlabeled studies (studies have not been reviewed yet), and black line is SVM decision plane. In (b) Weighting balances the training data by putting more weight on the minority class examples. In (c), aggressive undersampling balances the training data by throwing away majority class examples closest to the old decision plane in (a). When deciding which studies to be reviewed next, uncertainty sampling returns the unlabeled examples closest to the decision plane (U) while certainty sampling returns the unlabeled examples furthest to the decision plane from the bottom-left side (C).

By analyzing the differences between the state-of-the-art methods, we identified the following key components in solving the problem with active learning and linear SVM.

\vspace{3mm}
\noindent
{\em When to start training}: 
\begin{itemize}
\item \textbf{P} stands for ``patient''. As suggested by Wallace et al.~\cite{wallace2010semi}, ``hasty generation'', which means start training with too few relevant examples, may lead to poor performance. The algorithm keeps random sampling until a sufficient number of ``relevant'' studies are retrieved. In our experiments, the sufficient number of ``relevant'' studies retrieved is set to $5$, which means when at least $5$ ``relevant'' studies have been retrieved by random sampling, the algorithm goes into next stage. Wallace'10~\cite{wallace2010semi} and Miwa'14~\cite{miwa2014reducing} use \textbf{P} for when to start training.
\item
\textbf{H} stands for ``hasty'', which is the opposite of \textbf{P}. The algorithm starts training as soon as {\em ONE} ``relevant'' study is retrieved, as suggested in Cormack'14~\cite{cormack2014evaluation,cormack2015autonomy}.
\end{itemize}
{\em Which document to query next}: 
\begin{itemize}
\item
\textbf{U} stands for ``uncertainty sampling''. The algorithm utilizes uncertainty sampling to build the classifier, where unlabeled examples closest to the SVM decision plane are sampled for query (U in Figure~\ref{fig:SVM}). Wallace'10~\cite{wallace2010semi} uses \textbf{U} for Query strategy.
\item
\textbf{C} stands for ``certainty sampling''. The algorithm utilizes certainty sampling to build the classifier, where unlabeled examples furthest to the SVM decision plane and lie in the ``relevant'' side are sampled for query (C in Figure~\ref{fig:SVM}). Miwa'14~\cite{miwa2014reducing} and Cormack'14~\cite{cormack2014evaluation,cormack2015autonomy} use \textbf{C} for query strategy.
\end{itemize}
{\noindent\em Whether to stop training (or not)}: 
\begin{itemize}
\item
\textbf{S} stands for ``stop training''. The algorithm stops training
once the classifier is stable. In our experiments, the classifier is treated as stable once more than $30$ ``relevant'' studies have been retrieved as training examples. Wallace'10~\cite{wallace2010semi} uses \textbf{S} for whether to stop training.
\item
\textbf{T} stands for ``continue training''. The algorithm never stops
training as suggested in Cormack'14~\cite{cormack2014evaluation} and Miwa'14~\cite{miwa2014reducing}. If query strategy is \textbf{U}, algorithm switches to certainty sampling after classifier is stable but training never stops.
\end{itemize}
{\em How to balance the training data}: 
\begin{itemize}
\item
\textbf{N} stands for ``no data balancing''. The algorithm does not balance the training data (demonstrated in Figure~\ref{fig:nobal}) as suggested by Cormack'14~\cite{cormack2014evaluation}.
\item
\textbf{A} stands for ``aggressive undersampling''. The algorithm utilizes aggressive undersampling\footnote{Aggressive undersampling throws away majority (irrelevant) training examples closest to SVM decision plane until reaching the same number of minority (relevant) training examples. A demonstration is shown in Figure~\ref{fig:aggress}.} after classifier is stable, as suggested by Wallace'10~\cite{wallace2010semi}.
\item
\textbf{W} stands for ``Weighting''. The algorithm utilizes Weighting\footnote{Weighting assigns different weight to each class, $W_R = 1/|L_R|,\,W_I = 1/|L_I|$, when training SVM. A demonstration is shown in Figure~\ref{fig:weighting}. $L_R$ is defined in Figure~\ref{fig:Problem Statement}.} for data balancing (before and after the classifier is stable), as suggested by Miwa'14~\cite{miwa2014reducing}.
\item
\textbf{M} stands for ``mixing of Weighting and aggressive undersampling''. Weighting is applied before the classifier is stable while aggressive undersampling is applied after the classifier is stable. This treatment comes from the observation that ``Weighting'' performs better in early stages while ``aggressive undersampling'' performs better in later stages.
\end{itemize}
By combining different approaches, we ended up with 32 possible treatments including the state-of-the-art methods:
\begin{itemize}
\item
The \textbf{PUSA} approach advocated by Wallace'10~\cite{wallace2010semi}.
\item
The \textbf{PCTW} approach advocated by Miwa'14~\cite{miwa2014reducing}.
\item
The \textbf{HCTN} approach advocated by Cormack'14~\cite{cormack2014evaluation}.
\end{itemize}
Pseudo code for the 32 machine learning treatments are shown in Algorithm~\ref{alg:alg1}. Along with the current standard procedure as a baseline approach:
\begin{itemize}
\item
\textbf{Linear Review}: no machine learning, query studies in a random order.
\end{itemize}
All 32 machine learning treatments are tested and compared in Section~\ref{sect: Experiments}.

\noindent
\begin{minipage}[t]{.54\textwidth}
\vspace{0pt}
\centering
\begin{algorithm}[H]
\footnotesize
\SetKwInOut{Input}{Input}
\SetKwInOut{Output}{Output}
\SetKwInOut{Parameter}{Parameter}
\SetKwRepeat{Do}{do}{while}
\Input{$E$, set of all candidate studies\\$R$, set of ground truth relevant studies\\$AL$, algorithm code, e.g. \textbf{PUSA} for \\Wallace'10 treatment}
\Output{$L_R$, set of included studies}
\BlankLine

$L\leftarrow \emptyset$\; $L_R\leftarrow \emptyset$\; $\neg L\leftarrow E$\; 

\BlankLine
\tcp{Keep reviewing until stopping rule satisfied}
\While{$|L_R| < 0.95|R|$}{
    \tcp{Start training or not}
    \eIf{$|L_R| \geq Enough(AL)$}{
        \If{\textbf{T} $\in AL$ {\bf or} $NotStable(L_R)$}{
            $CL\leftarrow Train(AL,L)$\;
        }
        \tcp{Query next}
        $x\leftarrow Query(AL,CL,\neg L,L_R)$\;
    }{
        \tcp{Random Sampling}
        $x\leftarrow Random(\neg L)$\;
    }
    \tcp{Simulate review}
    $L\leftarrow L \cup x$\;
    $\neg L\leftarrow \neg L \setminus x$\;
    \If{$x\in R$}{
        $L_R\leftarrow L_R \cup x$\;
    }
}
\Return{$L_R$}\;
\caption{Psuedo Code}\label{alg:alg1}
\end{algorithm}
\end{minipage}\hfill
\begin{minipage}[t]{.46\textwidth}
\vspace{0pt}
\centering \small
~~~~~~~~\begin{tabular}{p{1.8in}}\hline
\rowcolor{gray!10}
The algorithm on this page use
the following notation. 
\begin{itemize}
\item
$E$: the set of all candidate studies (returned from search).
\item
$R\subset E$: the set of ground truth relevant studies.
\item
$I=E\setminus R$: the set of ground truth irrelevant studies.
\item
$L\subset E$: the set of labeled/reviewed studies, each review reveals whether a study $x\in R$, but incurs a cost.
\item
$\neg L=E\setminus L$: the set of unlabeled/unreviewed studies.
\item
$L_R=L\cap R$: the identified relevant (included) studies.
\item
$L_I=L\cap I$: the identified irrelevant (excluded) studies.
\end{itemize}
The general form of the excluding irrelevant studies problem can be described as following: start with $L=\emptyset$, prioritize which studies to be reviewed so as to maximize $|L_R|$ while minimizing $|L|$ (identify more relevant studies with less cost). 
  \end{tabular}
  \captionof{figure}{Notations and Problem Description}
  \label{fig:Problem Statement}
\end{minipage}
\SetAlgoFuncName{Support functions for Algorithm~\ref{alg:alg1}}{Function}
\begin{function}[H]
\footnotesize
$t1\leftarrow 5$\; $t2\leftarrow 30$\;
\BlankLine
\Fn{Enough($AL$)}{
    \uIf{\textbf{P} $\in AL$}{
        \Return{$t1$}\;
    }
    \uElseIf{\textbf{H} $\in AL$}{
        \Return{$1$}\;
    }
    \Else{
        \Return{$\infty$}\;
    }
}
\BlankLine
\Fn{NotStable($L_R$)}{
    \lIf{$|L_R|\geq t2$}{
        \Return{False}
    }
    \lElse{
        \Return{True}
    }
}
\BlankLine
\Fn{Train($AL,L$)}{
    \eIf{\textbf{W} $\in AL$ \textbf{or} \textbf{M} $\in AL$}{
        \tcp{Train linear SVM with Weighting}
        $CL\leftarrow SVM(L,kernel=linear,class\_weight=balanced)$\;
    }{
        \tcp{Train linear SVM}
        $CL\leftarrow SVM(L,kernel=linear)$\;
    }
    \If{\textbf{A} $\in AL$ \textbf{or} \textbf{M} $\in AL$}{
        \If{$\neg$ NotStable($L_R$)}{
            \tcp{Aggressive undersampling}
            $L_I\leftarrow L\setminus L_R$\;
            $tmp\leftarrow argsort(CL.decision\_function(L_I))[:|L_R|]$\;
            $CL\leftarrow SVM(L_R \cup tmp,kernel=linear)$\;
        }
    }
    \Return{$CL$}\;
}
\BlankLine
\Fn{Query($AL,CL,\neg L,L_R$)}{
    \eIf{\textbf{U} $\in AL$ \textbf{and} $NotStable(L_R)$}{
        \tcp{Uncertainty Sampling}
        $x\leftarrow argsort(abs(CL.decision\_function(\neg L)))[0]$\;
    }{
        \tcp{Certainty Sampling}
        $x\leftarrow argsort(CL.decision\_function(\neg L))[-1]$\;
    }
    \Return{$x$}\;
}
\caption{()}
\end{function}

\section{Experiments}
\label{sect: Experiments}

This section describes the experimental procedures that we used to evaluate the treatments described in Section~\ref{sect: Methods}. 

\subsection{Performance Metrics}
\label{sect: Performance Metrics}

As shown in Figure~\ref{fig:Problem Statement}, the problem to be solved is multi-objective. The performance of each algorithm is thus usually evaluated by its \textbf{recall ($|L_R|/|R|$)} vs. \textbf{studies reviewed ($|L|$)} curve. 

\noindent
This performance metrics is suggested by Cormack et al.~\cite{cormack2015autonomy,cormack2014evaluation,tredennick2015} and best fits the objectives of excluding irrelevant studies problem. To enable a statistical analysis of the performances, the \textbf{recall} vs. \textbf{studies reviewed} curve is cut by a $0.95$ \textbf{recall} line where \textbf{studies reviewed} ($|L|$) when reaching $0.95$ \textbf{recall} ($|L_R|\geq 0.95|R|$) is used to assess performances. The reason behind 
$0.95$ \textbf{recall} is that a) $1.00$ \textbf{recall} can never be guaranteed by any text mining method unless all the candidate studies are reviewed; b) $0.95$ \textbf{recall} is usually considered acceptable in evidence-based medicine~\cite{cohen2011performance,cohen2006reducing,o2015using} despite the fact that there might still be ``relevant'' studies missing~\cite{shemilt2016use}. As a result, two metrics are used for evaluation:
\begin{itemize}
\item
X95 = $\min \{|L| \mid |L_R|\geq0.95 |R|\}$.
\item
WSS@95 = $0.95-\text{X95}/|P|$.
\end{itemize}
Note that one algorithm is better than the other if its X95 is smaller or WSS@95~\cite{cohen2011performance} is larger.

\subsection{Datasets}
\label{sect: datasets}

Although a large number of SLRs are published every year, there is no dataset clearly documenting the details in primary study selection. As a result, three datasets are created reverse-engineering existing SLRs and being used in this study to simulate the process of excluding irrelevant studies. The three datasets are named after the authors of their original publication source-- Wahono dataset from Wahono et al. 2015~\cite{wahono2015systematic}, Hall dataset from Hall et al. 2012~\cite{hall2012systematic}, and Radjenovi{\'c} dataset from Radjenovi{\'c} et al. 2013~\cite{radjenovic2013software}. 

For each of the datasets, the search string \textbf{S} and the final inclusion list \textbf{F} from the original publication are used for the data collection. We retrieve the initial candidate collection \textbf{E} from IEEE Xplore with the search string (slightly modified to meet the requirement of IEEE Xplore). Then make a final list of inclusion \textbf{R} as \textbf{R} = \textbf{F} $\cap$ \textbf{E}. Here, for simplicity we only extract candidate studies from IEEE Xplore. We will explore possibilities for efficiently utilizing multiple data sources in the future work but in this paper, without loss of generality, we only extract initial candidate list from single data source. In this way, we created three datasets that reasonably resemble real SLR selection results assuming that any study outside the final inclusion list \textbf{F} is irrelevant to the original SLRs. A summary of the created datasets is presented in Table~\ref{tab: number}.

Apart from the three created datasets, one dataset (Kitchenham) is provided directly by the author of Kitchenham et al. 2010~\cite{kitchenham2010systematic} and includes two levels of relevance information. In general, only the ``content relevant'' labels are used in experiments for a fair comparison with other datasets. Additionally, the ``abstract relevant'' labels are used for detailed review cost analysis in RQ3. Summary of Kitchenham dataset is also presented in Table~\ref{tab: number}.

All the above datasets are available on-line at Seacraft, Zenodo\footnote{\em https://doi.org/10.5281/zenodo.1162952}.

\begin{table}
\caption{Descriptive statistics for experimental datasets}
\label{tab: number}
\begin{center}
\begin{tabular}{ |l|c|c|c|c| }
  \hline
   Datasets & \multicolumn{2}{|c|}{Generated} & \multicolumn{2}{|c|}{Original} \\
  \cline{2-5}
  & \#Candidate $|$\textbf{E}$|$ & \#Relevant $|$\textbf{R}$|$& \#Candidate & \#Relevant $|$\textbf{F}$|$\\
  \hline
  Wahono & 7002 & 62 & 2117 & 72\\
  \hline
  Hall & 8911 & 104 & 2073 & 136 \\
  \hline
  Radjenovi{\'c} & 6000 & 48 & 13126 & 106\\
  \hline
  Kitchenham & 1704 & 45 (132) & 1704 & 45 (132) \\
  \hline
\end{tabular}
\end{center}
{\footnotesize Our datasets are generated using information in the original SLR literature. Our candidate studies are retrieved by applying similar if not the same the search string from original SLR literature and search in IEEE Xplore. The set of our relevant studies is the intersection of the set of our candidate studies and the set of final included studies in the original SLR literature. Kitchenham dataset is different as it is provided directly by Kitchenham and it has two level of relevance labels-- 132 relevant studies by title and abstract review and within which, 45 relevant studies by content review.}
\end{table}

\subsection{Simulation Studies}
 
 In the following, each experiment is a simulation of one specific treatment on one dataset.
 More specifically,
there is no human activity involved in these experiments, when asked for a label, the true label in the dataset is queried instead of a human reviewer. As a result, each experiment can be repeated with different random seed to capture variances and also makes reproducing the experiments possible.

\subsection{Controlled Variables}
\label{sect: Controlled Variables}

For the sake of a fair comparison, different treatments in Section~\ref{sect: Methods} share an identical set of controlled variables including preprocessing, featurization and classifier. 

Each candidate study in the initial list is first tokenized by stop words removal after concatenating its title and abstract. After tokenization, the bag of words are featurized into a term frequency vector. Then, reduce the dimensionality of the term frequency vector with to keep only $M=4000$ of the terms with highest tf-idf\footnote{For term $t$ in document $d$, $Tfidf(t, d)=w^t_d\times (\log \frac{|D|}{\sum_{d\in D} sgn(w^t_d)}+1)$ where $w^t_i$ is the term frequency of term $t$ in document $d$. For term $t$, $Tfidf(t) = \sum_{d\in D} Tfidf(t,d) = \sum_{d\in D} w^t_d \times (\log \frac{|D|}{\sum_{d\in D} sgn(w^t_d)}+1)$ and is used for feature selection.} score and normalize the hashed matrix by its L2 norm each row at last. TfidfVectorizer in scikit-learn is utilized for the above preprocessing and featurization steps. Alternatives such as stemming, LDA~\cite{blei2003latent}, paragraph vectors~\cite{le2014distributed} require further exploration and are scheduled in our future works. All 32 treatments use the same classifier-- linear SVM from scikit-learn.

\section{Results}
\label{subsect: Results}

All the following results were generated from 30 repeats
simulations, using different random number seeds from each simulation.
As shown below, all our results
are reported in terms of medians (50th percentile) and iqrs ((75-25)th percentile).

\begin{table*}
\caption{Scott-Knott analysis for number of studies reviewed/ work saved over sampling to reach $95\%$ recall}
\label{tab: scottknott}

\begin{center}
\parbox{.49\linewidth}{
\centering
{\scriptsize \begin{tabular}{l@{~~~}l@{~~~}r@{~~~}r@{~~~}r@{~~~}r}
\arrayrulecolor{lightgray}
\multicolumn{2}{l}{\textbf{Wahono}}  & \multicolumn{2}{c}{\textbf{X95}} & \multicolumn{2}{c}{\textbf{WSS@95}}\\\hline
\textbf{Rank} & \textbf{Treatment} & \textbf{Median} & \textbf{IQR} & \textbf{Median} & \textbf{IQR} \\\hline
\rowcolor{green!40}
  1 &         HUTM &    670  &  230 & 0.85 & 0.04 \\
  1 &         HCTM &    740  &  220 & 0.84 & 0.03 \\
\hline  2 &         HUTA &    780  &  140 & 0.84 & 0.02 \\
  2 &         HCTW &    790  &  90 & 0.84 & 0.02 \\
  2 &         HUTW &    800  &  110 & 0.84 & 0.02 \\
  2 &         HCTA &    800  &  140 & 0.83 & 0.02 \\
\hline  3 &         PCTM &    1150  &  450 & 0.78 & 0.07  \\
  3 &         PUTM &    1180  &  420 & 0.78 & 0.07 \\
  3 &         PCTA &    1190  &  340 & 0.78 & 0.05 \\
  3 &         PUTA &    1190  &  340 & 0.78 & 0.05 \\
\rowcolor{red!30}
  3 &         PCTW &    1210  &  350 &  0.78 & 0.06 \\
  3 &         PUTW &    1220  &  370 &  0.77 & 0.06 \\
\hline  4 &         HUSM &    1410  &  400 & 0.75 & 0.06 \\
\hline  5 &         HUSA &    1610  &  370 & 0.72 & 0.07 \\
\hline  6 &         PUSM &    1810  &  370 & 0.69 & 0.06 \\
\rowcolor{red!30}
  6 &         PUSA &    1910  &  700 & 0.67 & 0.10 \\
\hline  7 &         HUSW &    2220  &  400 & 0.63 & 0.06 \\
  7 &         PUSW &    2240  &  360 & 0.63 & 0.06 \\
\hline  8 &         HUTN &    2700  &  40 & 0.56 & 0.01 \\
\rowcolor{red!30}
  8 &         HCTN &    2720  &  40 & 0.56 & 0.01 \\
  8 &         PCSW &    2860  &  1320 & 0.54 & 0.20 \\
  8 &         PCSM &    2860  &  1320 & 0.54 & 0.20 \\
  8 &         PCTN &    2850  &  1130 & 0.54 & 0.17 \\
  8 &         PUTN &    2850  &  1130 & 0.54 & 0.17 \\
\hline  9 &         PCSN &    3020  &  1810 & 0.51 & 0.26 \\
  9 &         PCSA &    3020  &  1810 & 0.51 & 0.26 \\
\hline 10 &         HUSN &    4320  &  110 &  0.33 & 0.03 \\
 10 &         PUSN &    4370  &  1290 & 0.32 & 0.19 \\
 \rowcolor{blue!50}
\hline 11 &       linear &    6650  &  0 & 0 & 0 \\
 11 &         HCSA &    6490  &  2760 & -0.01 & 0.39 \\
 11 &         HCSN &    6490  &  2760 & -0.01 & 0.39 \\
 11 &         HCSM &    6490  &  3110 & -0.01 & 0.44 \\
 11 &         HCSW &    6490  &  3110 & -0.01 & 0.44 \\
\hline \end{tabular}}
}
\parbox{.49\linewidth}{
\centering
{\scriptsize \begin{tabular}{l@{~~~}l@{~~~}r@{~~~}r@{~~~}r@{~~~}r}
\arrayrulecolor{lightgray}
\multicolumn{2}{l}{\textbf{Hall}}  & \multicolumn{2}{c}{\textbf{X95}} & \multicolumn{2}{c}{\textbf{WSS@95}}\\\hline
\textbf{Rank} & \textbf{Treatment} & \textbf{Median} & \textbf{IQR} & \textbf{Median} & \textbf{IQR} \\\hline
 1 &         HUTW &    340  &  90 & 0.91 & 0.01 \\
  1 &         HUTA &    340  &  130 & 0.91 & 0.02 \\
\rowcolor{green!40}
  1 &         HUTM &    350  &  120 & 0.91 & 0.01 \\
  1 &         HCTW &    370  &  60 & 0.91 & 0.01 \\
\hline  2 &         HUTN &    370  &  90 & 0.91 & 0.01 \\
  2 &         HCTM &    380  &  100 & 0.91 & 0.01 \\
  2 &         HCTA &    390  &  150 & 0.91 & 0.02 \\
   \rowcolor{red!30}
  2 &         HCTN &    410  &  80 & 0.90 & 0.01 \\
\hline  3 &         HUSM &    530  &  120 & 0.89 & 0.01 \\
  3 &         HUSW &    560  &  250 & 0.89 & 0.03 \\
   \rowcolor{red!30}
  3 &         PCTW &    610  &  210 & 0.88 & 0.02 \\
  3 &         PUTW &    610  &  220 & 0.88 & 0.03 \\
\hline  4 &         HUSA &    630  &  170 & 0.88 & 0.02 \\
  4 &         PCTN &    650  &  220 & 0.88 & 0.03 \\
  4 &         PUTN &    650  &  220 & 0.88 & 0.03 \\
  4 &         PUTM &    670  &  220 & 0.87 & 0.03 \\
  4 &         PCTM &    680  &  230 & 0.87 & 0.03 \\
  4 &         PCTA &    700  &  210 & 0.87 & 0.03 \\
  4 &         PUTA &    700  &  220 & 0.87 & 0.03 \\
  4 &         PUSW &    740  &  230 & 0.87 & 0.03 \\
\hline  5 &         PUSM &    770  &  240 & 0.86 & 0.03 \\
 \rowcolor{red!30}
  5 &         PUSA &    880  &  270 & 0.85 & 0.04 \\
\hline  6 &         PCSW &    1150  &  570 & 0.82 & 0.07 \\
  6 &         PCSM &    1150  &  570 & 0.82 & 0.07 \\
\hline  7 &         PCSN &    1530  &  1050 & 0.78 & 0.13 \\
  7 &         PCSA &    1530  &  1050 & 0.78 & 0.13 \\
  7 &         PUSN &    1550  &  1120 & 0.77 & 0.13 \\
  7 &         HUSN &    1800  &  1020 & 0.74 & 0.11 \\
\hline  8 &         HCSA &    7470  &  5980 & 0.03 & 0.67 \\
  8 &         HCSN &    7470  &  5980 & 0.03 & 0.67 \\
  \rowcolor{blue!50}
  8 &       linear &    8464  &  0 & 0 & 0 \\
  8 &         HCSM &    8840  &  6060 & -0.04 & 0.68 \\
  8 &         HCSW &    8840  &  6060 & -0.04 & 0.68 \\
\hline \end{tabular}}
}

\parbox{.49\linewidth}{
\centering
{\scriptsize \begin{tabular}{l@{~~~}l@{~~~}r@{~~~}r@{~~~}r@{~~~}r}
\arrayrulecolor{lightgray}
\multicolumn{2}{l}{\textbf{Radjenovi{\'c}}}  & \multicolumn{2}{c}{\textbf{X95}} & \multicolumn{2}{c}{\textbf{WSS@95}}\\\hline
\textbf{Rank} & \textbf{Treatment} & \textbf{Median} & \textbf{IQR} & \textbf{Median} & \textbf{IQR} \\\hline
\rowcolor{green!40}
  1 &         HUTM &    680   &  180  & 0.83 & 0.03 \\
  1 &         HCTM &    780   &  130  & 0.82 & 0.02 \\
  1 &         HCTA &    790   &  180  & 0.82 & 0.03 \\
  1 &         HUTA &    800   &  180  & 0.82 & 0.03 \\
\hline  2 &         HUSA &    890   &  310  & 0.80 & 0.06 \\
  2 &         HUSM &    890   &  270  & 0.80 & 0.05 \\
\hline  3 &         HUTW &    960   &  80  & 0.79 & 0.02 \\
  3 &         HCTW &    980   &  60  & 0.79 & 0.01 \\
  3 &         HUSW &    1080   &  410  & 0.77 & 0.07 \\
\hline  4 &         PCTM &    1150   &  270  & 0.76 & 0.05 \\
  4 &         PUTM &    1150   &  270  & 0.76 & 0.05 \\
\hline  5 &         HUTN &    1250   &  100  &  0.74 & 0.02 \\
  5 &         PCTA &    1260   &  210  & 0.74 & 0.05 \\
  5 &         PUTA &    1260   &  210  & 0.74 & 0.05 \\
  \rowcolor{red!30}
  5 &         HCTN &    1270   &  70  & 0.74 & 0.02 \\
  5 &         PUSM &    1250   &  400  & 0.74 & 0.07 \\
  5 &         PUSW &    1250   &  450  & 0.73 & 0.08 \\
  5 &         PUTW &    1350   &  310  & 0.72 & 0.06 \\
\rowcolor{red!30}
  5 &         PCTW &    1370   &  310  & 0.72 & 0.06 \\
  \rowcolor{red!30}
  5 &         PUSA &    1400   &  490  & 0.71 & 0.09 \\
\hline  6 &         HUSN &    1570   &  300  & 0.69 & 0.05 \\
  6 &         PCTN &    1600   &  360  & 0.68 & 0.06 \\
  6 &         PUTN &    1600   &  360  & 0.68 & 0.06 \\
\hline  7 &         PUSN &    1890   &  320  &  0.64 & 0.06 \\
\hline
  8 &         PCSW &    2250   &  940  & 0.57 & 0.20 \\
  8 &         PCSM &    2250   &  940  & 0.57 & 0.20 \\
\hline  9 &         PCSN &    2840   &  1680  & 0.47 & 0.31 \\
  9 &         PCSA &    2840   &  1680  & 0.47 & 0.31 \\
\hline 10 &         HCSA &    5310   &  2140  & 0.07 & 0.36 \\
 10 &         HCSN &    5310   &  2140  & 0.07 & 0.36  \\
 10 &         HCSM &    5320   &  2200  &  0.02 & 0.37  \\
 10 &         HCSW &    5320   &  2200  & 0.02 & 0.37 \\
 \rowcolor{blue!50}
 10 &       linear &    5700   &  0  & 0 & 0 \\
\hline \end{tabular}}}
\parbox{.49\linewidth}{
\centering
{\scriptsize \begin{tabular}{l@{~~~}l@{~~~}r@{~~~}r@{~~~}r@{~~~}r}
\arrayrulecolor{lightgray}
\multicolumn{2}{l}{\textbf{Kitchenham}}  & \multicolumn{2}{c}{\textbf{X95}} & \multicolumn{2}{c}{\textbf{WSS@95}}\\\hline
\textbf{Rank} & \textbf{Treatment} & \textbf{Median} & \textbf{IQR} & \textbf{Median} & \textbf{IQR} \\\hline
  1 &         HUSA &    590  &  170 & 0.60 & 0.19 \\
  1 &         HUTA &    590  &  80 & 0.60 & 0.06
 \\
  1 &         HUSM &    620  &  70 & 0.58 & 0.04
 \\
 \rowcolor{green!40}
  1 &         HUTM &    630  &  110 & 0.58 & 0.07 \\
  \rowcolor{red!30}
  1 &         PUSA &    640  &  130 & 0.57 & 0.08 \\
  1 &         HUSW &    640  &  140 & 0.57 & 0.09 \\
\hline  2 &         HUTN &    680  &  30 & 0.55 & 0.02 \\
  2 &         HCTA &    680  &  100 & 0.55 & 0.08 \\
  2 &         PUSM &    680  &  90 & 0.55 & 0.06
 \\
  2 &         HCTM &    680  &  110 & 0.55 & 0.07 \\
  2 &         PCTM &    690  &  90 & 0.54 & 0.06 \\
  2 &         PUTM &    690  &  70 & 0.54 & 0.05 \\
  2 &         PUTA &    710  &  110 & 0.53 & 0.08 \\
  2 &         HUTW &    710  &  20 & 0.53 & 0.02 \\
\hline  3 &         PUSW &    720  &  110 & 0.52 & 0.08 \\
  3 &         PCTA &    720  &  100 & 0.52 & 0.08 \\
  \rowcolor{red!30}
  3 &         HCTN &    730  &  60 & 0.52 & 0.04 \\
  3 &         HCTW &    750  &  60 & 0.51 & 0.04 \\
  3 &         PUTN &    750  &  80 & 0.51 & 0.05 \\
\hline  4 &         PCTN &    750  &  80 & 0.51 & 0.05 \\
  4 &         PUTW &    780  &  70 & 0.49 & 0.04 \\
  \rowcolor{red!30}
  4 &         PCTW &    780  &  150 & 0.49 & 0.09 \\
\hline  5 &         PUSN &    800  &  140 & 0.47 & 0.09 \\
  5 &         HUSN &    870  &  280 & 0.43 & 0.16\\
\hline  6 &         PCSW &    990  &  330 & 0.35 & 0.19 \\
  6 &         PCSM &    990  &  330 & 0.35 & 0.19 \\
  6 &         PCSN &    1050  &  370 & 0.32 & 0.24 \\
  6 &         PCSA &    1050  &  370 & 0.32 & 0.24 \\
   \rowcolor{blue!50}
\hline  7 &       linear &    1615  &  0 & 0 & 0 \\
  7 &         HCSA &    1670  &  60 & -0.04 & 0.04 \\
  7 &         HCSN &    1670  &  60 & -0.04 & 0.04 \\
  7 &         HCSM &    1680  &  60 & -0.04 & 0.04 \\
  7 &         HCSW &    1680  &  60 & -0.04 & 0.04 \\
\hline \end{tabular}}}

\end{center}
{\footnotesize Simulations are repeated for $30$ times, medians ($50$th percentile) and iqrs (($75$-$25$)th percentile) are presented. Smaller/larger median value for X95/WSS@95 represents better performance while smaller iqr means better stability. Treatments with same rank have no significant difference in performance while treatments of smaller number in rank are significantly better than those of larger number in rank. The recommended treatment FASTREAD is colored in {\setlength{\fboxsep}{1pt}\colorbox{green!40}{green}} while the state-of-the-art treatments are colored in {\setlength{\fboxsep}{1pt}\colorbox{red!30}{red}} and linear review is colored in {\setlength{\fboxsep}{1pt}\colorbox{blue!50}{blue}}.}

\end{table*}

\newpage
{\bf RQ1: Can active learning techniques reduce effort in primary study selection?} 

In Table~\ref{tab: scottknott}, we tested 32 active learning treatments and linear review. According to the results, most active learning treatments perform consistently better than linear review (colored in {\setlength{\fboxsep}{1pt}\colorbox{blue!50}{blue}}) on all four datasets while four treatments (\textbf{HCS*}) can be even worse than linear review. Interestingly these four treatments share same codes of \textbf{HCS}, which hastily start training (\textbf{H}) with greedy query strategy (\textbf{C}) and give up the attempt to correct the model short after (\textbf{S}). The problem of ``hasty generation'' is maximized in the setting of \textbf{HCS} and thus leads to an even worse performance than linear review. In general, other active learning treatments can reduce review costs by allowing the reviewer to read fewer studies while still find 95\% of the relevant ones. As for how much effort can be saved, \textbf{RQ3} will answer the question in details.

Based on the above, we say:
\begin{lesson}
    In general, active learning techniques can reduce cost in primary study selections with a sacrifice of (say 5\%) recall.
\end{lesson}

{\bf RQ2: Should we just adopt the state-of-the-art treatments from other fields? Is it possible to build a better one by mixing and matching from those?}

In Table~\ref{tab: scottknott},  performance of the three state-of-the-art treatments are colored in {\setlength{\fboxsep}{1pt}\colorbox{red!30}{red}}. On Wahono datasets, Miwa'14 (\textbf{PCTW}) outperforms the other two treatments; while on Hall dataset, Cormack'14 (\textbf{HCTN}) has the best performance; on Radjenovi{\'c} dataset, all three treatments perform similarly; and on Kitchenham dataset, Wallace'10 (\textbf{PUSA}) outperforms the others. Neither of the three state-of-the-art treatments consistently performs the best. This means that adopting the state-of-the-art treatments will not produce best results. According to Scott-Knott analysis, the performance of one treatment, \textbf{HUTM} (colored in {\setlength{\fboxsep}{1pt}\colorbox{green!40}{green}}), consistently stays in the top rank across all four datasets.
Further, this treatment dramatically out-performs
all three state-of-the-art treatments by requiring 20-50\% fewer studies to be reviewed to reach 95\% recall.
We call this treatment FASTREAD. It executes as follows:
\begin{enumerate}
\item
Randomly sample from unlabeled candidate studies until 1 ``relevant'' example retrieved.
\item
Then start training with weighting and query with uncertainty sampling, until 30 ``relevant'' examples retrieved.
\item
Then train with aggressive undersampling and query with certainty sampling until finished.
\end{enumerate}
 
 Hence, our answer to this research question is:

\begin{lesson}
    No, we should not just adopt the state-of-the-art methods from other fields. A better method called FASTREAD is generated by mixing and matching from the state-of-the-art methods.
\end{lesson}

\textbf{RQ3: How much effort can FASTREAD save in an SLR?}

In terms of the number of studies reviewed, WSS@95 scores in Table~\ref{tab: scottknott} reflects how much FASTREAD can save. Number of ``relevant'' studies ($|R|$) and the total number of candidate studies ($|C|$) affect WSS@95 a lot, e.g. WSS@95=0.58 in Kitchenham dataset with $|R|$=45, $|C|$=1704 and WSS@95=0.91 in Hall dataset with $|R|$=104, $|C|$=8911. Even the smallest number of WSS@95=0.58 in Kitchenham dataset is a success in the reduction of number of studies need to be reviewed comparing to the 5\% recall lost.

The above performance metrics can be used for comparing the performance of different algorithms. However, for a more realistic cost analysis, labeling/reviewing each study has different costs. For each studies in $L$, its abstract and title has been reviewed, thus costs $C_A$. In addition, there exists a set $L_D\subset L,\,L_R\subset L_D$ where studies in $L_D$ have been reviewed by their contents, thus cost an additional $C_D$ for each study.  Table~\ref{tab: save} shows how much FASTREAD save over reviewing all candidate studies. Suppose $C_D=9C_A$, following the estimation that Shemilt made: 1 minute to screen a title-abstract record, 4 minutes to retrieve a full-text study report, and 5 minutes to screen a full-text study report~\cite{shemilt2016use}. Then the reduction in review cost is $(32C_D+1074C_A)/(132C_D+1704C_A) = 47.1\%$\footnote{According to Table~\ref{tab: number}, reviewing all studies costs $132C_D+1704C_A$. In our simulations, in average FASTREAD did 630 abstract reviews and 100 content reviews.}. On other datasets, although we do not have the exact number of ``abstract relevant'' studies, we can estimate the worst case review cost reduction\footnote{In the worst case we assume that every study reviewed is ``abstract relevant'' and thus costs $C_D+C_A$ to review and there is no ``abstract relevant'' study left except for the 5\% missing ``content relevant'' ones. E.g. in Wahono dataset, FASTREAD reviews 670 studies among the 7002 candidate ones, it costs $670(C_A+C_D)$ while reviewing all studies costs $(670+4)C_D+7002C_A$.} with the numbers in Table~\ref{tab: number} and Table~\ref{tab: save}: a) Wahono dataset: $1-670(C_A+C_D)/((670+4)C_D+7002C_A) = 48.7\%$; b) Hall dataset: $1-350(C_A+C_D)/((350+6)C_D+8991C_A) = 71.3\%$; c) Radjenovi{\'c} dataset: $1-680(C_A+C_D)/((680+3)C_D+6000C_A) = 44.0\%$. 
Note that training time costs are negligibly small (1 second for each round in average) compared to the review time $C_A$ because of the small training size (less than 1000 examples before reaching 95\% recall). 

\begin{lesson}
    Our results and estimations suggest that FASTREAD can save 
   more than 40\% of the effort (associated with the primary selection study
phase of a literature review)
 while retrieving $95\%$ of the ``relevant'' studies.
\end{lesson}


\begin{table}
\caption{How much can FASTREAD save?}
\label{tab: save}
\begin{center}
\begin{tabular}{ |l|c|c|c|}
  \hline
   Datasets & \# Studies Reviewed & Review Cost & \# Missing Relevant \\
  \hline
  Wahono & $7002-670=6332$ & $\geq6332C_A+4C_D$ & 4\\
  \hline
  Hall & $8991-350=8641$ & $\geq8641C_A+6C_D$ & 6\\
  \hline
  Radjenovi{\'c} & $6000-680=5320$ & $\geq5320C_A+3C_D$ & 3\\
  \hline
  Kitchenham & $1704-630=1074$ & $32C_D+1074C_A$ & 3\\
  \hline
\end{tabular}
\end{center}
{\footnotesize Numbers of reviewing every candidate study minus numbers of reviewing with FASTREAD. For example, on Kitchenham dataset, FASTREAD reviews 1074 fewer studies, which costs $32C_D+1074C_A$ less review effort, while misses 3 ``relevant'' ones. Here $C_D$ is the cost to review a study by its content and $C_A$ is the cost to review a study by its title and abstract.}
\end{table}

\section{Tool Support}
\label{sect: tool}

\begin{figure}[!h]
    \centering
    \includegraphics[width=\linewidth]{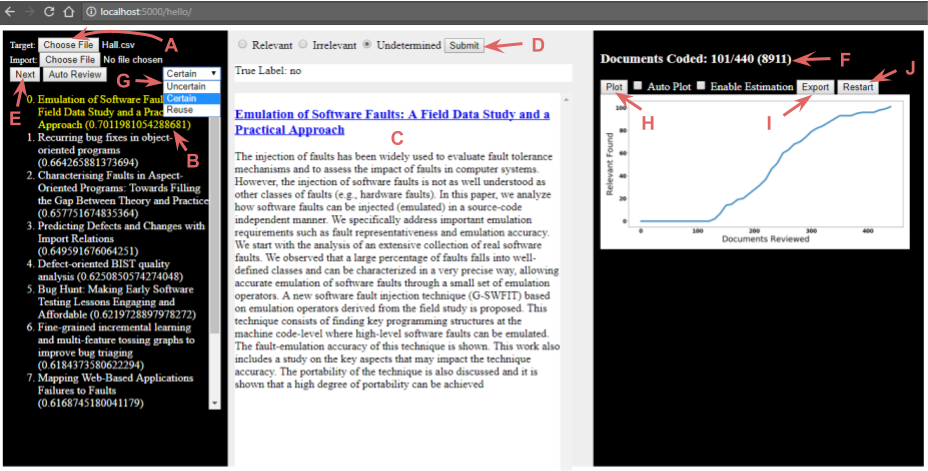}
    \caption{Basic interface of the FASTREAD tool.}
    \label{fig:FASTREAD}
\end{figure}

In order to implement FASTREAD, we developed a simple tool as shown in Fig.~\ref{fig:FASTREAD}. This software is freely available from SeaCraft Zenodo at \textit{https://doi.org/10.5281/zenodo.837861} and its Github repository at \textit{https://github.com/fastread/src}.

\begin{figure}[!h]
    \centering
    \subfloat[Input format]
    {
        \includegraphics[width=0.48\linewidth]{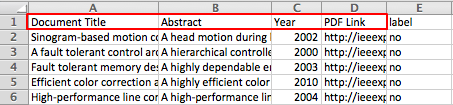}\label{fig:input}
    }
    \subfloat[Output format]
    {
        \includegraphics[width=0.48\linewidth]{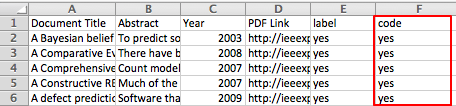}\label{fig:output}
    }    
    
    \caption{Data format for FASTREAD tool.}
    \label{fig:csv}
\end{figure}

Using FASTREAD, a review starts with \textbf{A}: selecting the input candidate study list from \textit{workspace/data/} directory. The input candidate list is specified in the format shown in Fig.~\ref{fig:input}. The input CSV file must have the \textit{Document Title}, \textit{Abstract}, \textit{Year}, and \textit{PDF Link} columns. The \textit{label} column, which is the true label of the candidate studies, is optional and is only used for testing. The output CSV file generated by the FASTREAD tool has an additional \textit{code} column, which is the reviewer-decided label for the candidate study. The final inclusion list can be retrieved by extracting all the studies with ``yes'' in the \textit{code} column.

 \noindent As shown by the annotations in Fig.~\ref{fig:FASTREAD},
 reviews using FASTREAD    proceeds as follows:
\begin{enumerate}
\item[\textbf{B}] Randomly select $10$ candidate studies for review.
\item[\textbf{C}] Read through the title and abstract (and click on the title and read the full text if needed) of the candidate study.
\item[\textbf{D}] Decide whether this study should be coded as \textit{Relevant} or \textit{Irrelevant} then click   \textit{Submit}.
\item[\textbf{E}] Click the \textit{Next} button to save   codes.   $10$ more
candidates are then   selected.
\item[\textbf{F}] The review status will change every time new studies are coded by reviewer and the \textit{Next} button is hit. The status is shown in the format ``Documents Coded: \textit{Number of relevant studies found} / \textit{Number of studies reviewed} (\textit{Total number of candidate studies}).''
\item[\textbf{G1}] Once \textbf{1} ``relevant'' study is coded, \textit{Random sampling} moves to  \textit{Uncertainty sampling}.
\item[\textbf{G2}] Once \textbf{30} ``relevant'' study is coded, \textit{Uncertainty sampling} can   change \textit{Certainty sampling}.
\item[\textbf{H}] Fig. H can be plotted by clicking the \textit{Plot} button or checking \textit{Auto Plot} (figure cached in   \textit{src/static/image/} directory).
\item[\textbf{I}] Once finished, coded studies can be exported into a CSV file in the \textit{workspace/coded/} directory, in the format shown in Fig.~\ref{fig:output}.
\end{enumerate}
Note that the \textit{Restart} button (\textbf{J}) is only for testing and discards all codes.


\section{Discussion}
\label{sect: Discussions}

\subsection{What is missed?}

Our results will show, with FASTREAD, $95\%$ of the ``relevant'' studies can be retrieved by reviewing a small portion (usually hundreds of studies) of long candidate study list. Given that, it is wise to reflect
on the 5\% of papers {\em not} found by such an analysis. To this end, we took one of our case studies and reflected on:
\begin{itemize}
\item The set of papers $R$ that a human analyst declared to be ``relevant'' (as listed in their reference list at the end of their paper).
\item The {\em tangentially relevant} subset of those  papers $R_1 \subseteq R$ that a human analyst explicitly mentions, however briefly, in the body of their paper.
\item The yet smaller subset of those papers $R_2 \subseteq R_1$  that a human analyst discusses, at length, in the body of their report (and for
our purposes ``at length'' is ``more that two lines''). We call these {\em insightful papers}. Clearly, FASTREAD should not be recommended if our method always misses the insightful papers. 
\end{itemize}
For our case studies, on 30 repeats of our methods, we found that $R_2\setminus L_R=\emptyset$; i.e. FASTREAD never missed an insightful paper. As for the tangentially
relevant papers, FASTREAD found all of those in 95\% of the 30 repeats. 
Based on this analysis, we infer that missing  $5\%$ of the papers is not a major impediment to using FASTREAD. Similar conclusion was derived by Shemilt et al. in 2016~\cite{shemilt2016use}. More interestingly, we found that more than 90\% of the missing studies come from a same set of size of $0.1|R|$. Which means some studies are more likely to be missed while most studies have never been missed in 30 repeats. This may suggest that there are outliers in relevant studies, which are not very important according to our case studies.

That said,  if the SLR conductor does not want to miss any potential relevant study,  they need to review all the candidate studies with full cost. We are actively exploring possibilities to mitigate or compensate the missing studies issue. 
For example, one technique is ``diversity sampling''; i.e. to explore unknown regions by sampling the least similar studies from what have been reviewed before. Exploration and exploitation can be balanced by selection different weight between diversity sampling and certainty/uncertainty sampling Note that more exploration  means fewer missing studies but higher review cost.



\subsection{What about domain knowledge?}

In our simulations, we assume that no initial seed training set is available thus a random sampling is performed to collect the minimum training set. This assumption represents the worst case while no external knowledge is available. We show in this work that the absence of that domain knowledge is not a critical failing of the approach. On the other hand, such domain knowledge usually exists in real world SLRs and will boost the performance of FASTREAD if wisely used. For example, if one relevant example and one irrelevant example are known in the very beginning, the random sampling step of FASTREAD is no longer needed and thus leads to additional cost reduction. More details about how to wisely use domain knowledge to boost FASTREAD will be explored further after this work. While we have some preliminary results in that area, we have nothing definitive to report at this time.

\subsection{What about real human reviewers?}

In our simulations, we assume that there is only one reviewer who never make mistakes. In real world SLRs, there will be multiple reviewers who make some mistakes. 

To handle this,
 FASTREAD could be changed to one central learner with multiple review agents. Every agent reviews different studies and feedback his or her decisions to the central learner. The central learner then trains on the feedback of every agent and assigns studies to each agent for review. Such schema will keep all the property of single reviewer FASTREAD and performs similarly. In addition, there might be more intelligent way to allocate review tasks based on the different performance of review agents~\cite{wallace2011should}.

Second, consider those multiple reviewers now make mistakes. Candidate studies need to be reviewed by multiple reviewers in case any of them makes mistakes. To explore this issue, appropriate data need to be collected on how human reviewers make mistakes. Wallace et al. addressed this issue in 2015~\cite{nguyen2015combining} by analyzing the best policy for allocating review tasks to reviewers with different experience levels as well as difference costs. We also plan to to address this issue in our future work.

\subsection{What about multiple categories of studies?}

In our simulations, we assume that the target is binary classification. However, primary study selection in real world SLRs might be a multi-label classification problem. For example, an SLR with two research questions might go through a primary study selection while each candidate is labeled as ``relevant to RQ1'', ``relevant to RQ2'', or ``irrelevant'' while the first two labels can co-exist. The simplest solution for this is to run multiple FASTREAD learners each learns on one label vs. others and each reviewer classify on one label only. In this case, the multi-label classification problem can be divided into multiple FASTREAD problems. Additional work such as ensemble learners can be explored in future works.

\section{Threats to Validity}
\label{sect: Threats to Validity}

There are several validity threats to the design of this study~\cite{feldt2010validity}. Any conclusions made from this work must be considered with the following issues in mind:

{\em Conclusion validity} focuses on the significance of the treatment. To
enhance the conclusion validity of this work, we employed several statistical
tests (Scott-Knott) to reduce the changes of making spurious conclusions. 

{\em Internal validity} measures  how sure we can be that the treatment
actually caused the outcome. To enhance   internal validity, we heavily constrained our experiments
(see our simulated results in strictly controlled environments as discussed in Section~\ref{sect: Controlled Variables}).

{\em Construct validity} focuses on the relation between the theory
behind the experiment and the observation. In this work, we evaluated
our results via different treatments with WSS@95 as stated in Section~\ref{sect: Performance Metrics}-- note that those
measures took us as close as we can to computing
cost reduction without ``abstract relevant'' information. 
That is, it fits the objective of human-in-the-loop primary study selection as defined in the current literature~\cite{tredennick2015,cormack2015autonomy,cormack2014evaluation}. Increasing the number of different measures may increase construct validity
so, in future work, we will further explore more metrics.

{\em External validity }concerns how well the conclusion can be applied outside. All the conclusions in this study are drawn from the experiments running on three software engineering SLR datasets created with information from Hall, Wahono, Radjenovi{\'c} et al. studies~\cite{hall2012systematic,wahono2015systematic,radjenovic2013software} and one dataset provided by Kitchenham~\cite{kitchenham2010systematic}. Therefore, such conclusions may not be applicable to datasets of different scenarios, e.g., citation screening from evidence based medicine or TAR from e-discovery. Such bias threatens any classification experiment. The best any researcher can do is to document that bias then make available to the general research community all the materials used in a study (with the hope that other researchers will explore similar work on different datasets). Existing active learning techniques in citation screening have been criticized by Olorisade et al. for being not replicable~\cite{olorisade2016critical,olorisade2017reproducibility}. To this end, we have published all our code at \textit{https://github.com/fastread/src} and all our data at \textit{https://doi.org/10.5281/zenodo.1162952}.

In the experiments, we assume that the human reviewer is always correct. In practice, this assumption cannot hold and problems such as disagreement between reviewers or concept drift (in which reviewers disagree with themselves as time passes) may occur.  As discussed
below when we discuss {\em Future Work}, we intend to explore this matter in the near future.

The comparisons in our experiment are based on the controlled variables listed in Section~\ref{sect: Controlled Variables}. If those
settings change,
then the conclusion in Section~\ref{subsect: Results} may become unreliable.

\section{Conclusions}\label{sect: Conclusion}

Systematic literature reviews are the primary method for aggregating evidence in evidence-based software engineering. It is suggested for every researcher in software engineering to frequently conduct SLRs~\cite{keele2007guidelines}.
One drawback with such SLRs is the time
required to complete such a study:
  an SLR would can weeks to  months to finish and the conclusion drawn can be out of date in a few years. 
  
  To tackle this barrier to understanding the literature, this study focuses on primary study selection, one of the most difficult and time consuming steps in an SLR. Machine learning methods, especially active learning, are explored in our attempts to reduce the effort required to exclude primary studies. In this paper:
  
\begin{itemize}
\item
We explored 32 different active learners. To the best of our knowledge, this is largest
such study yet completed in the software engineering domain. 
\item
We  have collected data from four large literature reviews. This data is publically available (doi.org/10.5281/zenodo.1162952). Note that the creation and distribution of these data sets is an important contribution, because prior to this study, it was  difficult to obtain even one such data set. 
\item
We have offered a baseline result that can serve
as a challenge problem for SE researchers: how to find more relevant
papers after reviewing fewer  papers. 
We  have  placed in the public domain (github.com/fastread/src) software
tools that let others compare our approach with alternative methods.
\item
We created a new reading-assistant tool called FASTREAD.
To the best of our knowledge, FASTREAD's  combination of methods  has  not been previously explored.
\item
Using FASTREAD, we decreased the number of studies to be reviewed by 20-50\% (comparing to the prior state-of-the-art).
\end{itemize}
As a result  of the above we can:
\begin{itemize}
\item Offer much assistance to any future SLR.
\item Offer a cautionary tale to SE 
researchers who use data miners. Specifically: do not be content with off-the-shelf solutions developed
by other communities. SE   has nuanced differences to other domains so our methods need to be tuned to our data. Even within the SE community there may be variations, so the framework provided by this paper is a good example to find the best method for a specific task on specific data.
\end{itemize}

\section{Future Work}
This study has several limitations as described in Section~\ref{sect: Discussions}. We consider the limitations as open challenges and plan to address those in future work. Specific problems and plans for the future are listed below.

\begin{itemize}

\item
{\em Conclusions are drawn from three synthetic SLR datasets and one Kitchenham dataset.} Validate the generalizability of the results on different datasets, including datasets from evidence-based medicine and e-discovery.

\item
{\em Experiment results are evaluated by WSS@95, which assumes a stop rule of reaching 95\% recall.} How to stop at 95\% recall without first knowing the number ``relevant'' studies in the pool is an interesting topic. We are exploring this topic actively.

\item
{\em The size and prevalence of data can affect performance of FASTREAD.} With the capability of cost reduction from FASTREAD, it is reasonable to ask whether we need the narrow initial search. An interesting future research would be to use every paper on, say Scopus, database as candidates and allow user to just using some simple search to initiate and guide the selection. As a result, the recall is no longer restricted by the initial search string thus may yield higher recall with reasonable cost.

\item
{\em About $10\%$ to $20\%$ efforts are spent on random selection step and most of the variances are also introduced in this step.} To speed up the random selection step, external expert knowledge will be introduced while unsupervised learning methods such as VTM, LDA, word2vec, or t-SNE will also be considered in future work. 

\item
{\em Some magic parameters are arbitrarily chosen, which may affect the performance.} However, parameter tuning is not a good fit for human-in-the-loop primary study selection because a) parameters should be tuned for the data working on; b) but the effect of applying different parameters can not be tested since querying extra label incurs extra cost. Therefore, novel methods should be explored for parameter selection; e.g. better criterion for when to switch from uncertainty sampling to certainty sampling (instead of the ``30'' relevant examples rule applied now). Works from Borg~\cite{Borg2016TuneR} and Fu~\cite{Fu2016Tuning} will be considered as candidate solutions to this problem.

\item
{\em Current scenario is restricted to having only one reviewer, which is impractical in practice.} Problems including how to assign review tasks to multiple reviewers and how to utilize reviewers with different cost and different capability will be explored in the future.

\item
{\em Currently, we assume  that  reviewers never make mistakes.} In future work,
we will explore  concept drift (reviewers disagree with themselves, at some later time) and how to settle disagreements (reviewers disagree with each other).

\item
{\em This study focuses only on primary study selection.} Assistance on other steps of SLR such as searching, data extraction, and protocol development can also help reduce total effort of SLRs. The potential of combining VTM, snowballing, and other tools with FASTREAD needs to be explored as well.

\end{itemize} We invite other researchers to join us in the exploring the above. 


\section*{Acknowledgement}
The authors thank Barbara Kitchenham for
her attention to this work and for
sharing with us the ``Kitchenham'' dataset used in our experiments.

\bibliographystyle{spmpsci}

\end{document}